\documentclass[11pt,twoside]{article}
\usepackage{asp2010}

\resetcounters

\begin{document}

\title{The QuickReduce data reduction pipeline for the WIYN One Degree Imager}
\author{Ralf Kotulla
\affil{Center for Gravitation, Cosmology and Astrophysics, \\ Department of
  Physics, University of Wisconsin - Milwaukee, \\ 1900 E Kenwood Blvd, Milwaukee,
  WI 53219, USA}
}

\begin{abstract}
Optimizing one's observing strategy while at the telescope relies on knowing
the current observing conditions and the obtained data quality. In particular
the latter is not straight forward with current wide-field imagers, such as the
WIYN One Degree Imager (ODI), currently consisting of 13 detectors, each of them
read out in 64 independent cells.  

Here we present a fast data reduction software for ODI, optimized for a first
data inspection during acquisition at the the telescope, but capable enough for
science-quality data reductions. The pipeline is coded in pure python with
minimal additional requirements. It is installed on the ODI observer's interface
and publicly available from the author's webpage. It performs all basic
reduction steps as well as more advanced corrections for pupil-ghost removal,
fringe correction and masking of persistent pixels. Additional capabilities
include adding an accurate astrometric WCS solution based on the 2MASS reference
system as well as photometric zeropoint calibration for frames covered by the
SDSS foot-print.

The pipeline makes use of multiple CPU-cores wherever possible, resulting in an execution
time of only a few seconds per frame.  As such this QuickReduce
pipeline offers the ODI observer a convenient way to closely monitor data
quality, a necessity to optimize the observing strategy during the night.

\end{abstract}

\section{The challenge of observing with modern multi-detector imagers}

Astronomy in particular in recent times is a largely data-driven science, and
ground-breaking discoveries are often made with the newest instruments. While
imaging the sky is hardly new, modern wide-field imagers offer larger sky
coverage at better sensitivities, allowing us to probe deeper into the history
of the universe. However, choosing the optimal observing strategy requires to
keep a close eye on the current sky-conditions, a highly non-trivial task when
faced with multiple CCD detectors being read out in different channels, each
with its own characteristics. Having real-time access to data that is reduced
far enough to have most instrument signatures removed is hence critical for
guiding the observing program and obtaining the best possible data.

\section{WIYN and the One Degree Imager}
The One Degree Imager is the new flagship camera on WIYN, the Wisconsin-Indiana-
Yale-NOAO telescope, located at the National Observatory on Kitt Peak near
Tucson, Arizona \citep{Harbeck2010}. Its current version, partial ODI or pODI,
consists of 13 Orthogonal Transfer Arrays, each consisting of 64 imaging cells
of 490x484 pixels. Each sub-array is read out independently from all others and
thus has its own detector characteristics, in particular overscan level and
gain. In total, each exposure results in 832 independent images. To address
difficulties associated with handling ODI's large data volumes, WIYN has devised
the ODI Portal, Pipeline, and Archive (PPA). The pipeline described here is
implemented into PPA and allows for user-specified data reduction capabilities
in the cloud - without having to download the raw data first.

\section{QuickReduce: Algorithm and Features}
The QuickReduce (QR) pipeline was originally conceived as a tool to support
commissioning of ODI to assess data quality while observing. Starting at that
time it quickly became a test-bed to rapidly prototype corrections and develop
specific reduction recipes, as well as to cross-check the Automatic Calibration
pipeline (AuCaP), an IRAF-based pipeline developed by NOAO. At the time of
writing, QR performs the following reduction steps: As a first step before any
corrections are done, QR creates a catalog of saturated pixels to later mask
pixels affected by saturation and/or persistency effects. The basic instrument
de-trending consists of (in order of execution) crosstalk-correction, overscan
subtraction, bias, and dark subtraction, followed by correction for
non-linearity on a cell-by-cell basis, and finally flatfield correction. During
this reduction, all 64 cells of any given OTA are combined into one monolithic
4K$\times$4K image.

After primary detrending QR uses the catalogs of saturated stars of the present
frame to mask out all pixels in columns above saturated pixels as these are
affected by charge smearing due to issues with the charge transfer efficiency of
the current detector generation. Furthermore, saturated stars leave downward
trails that become fainter with time; these are masked out using catalogs of
saturated stars in earlier frames. 

Fringing, caused by interference of night sky emission lines in the thin
detector material, affects observations in the i' and z' bands. To minimize
fringing in the exposure, QR uses static fringe maps generated from long
exposures taken under dark sky conditions. While the structure of the fringing
does not vary with time, the amplitude of this effect is strongly
time-variable. To derive fringe scaling factors, QR implements the method
described in \cite{Snodgrass2013} that works both fast and reliable in all cases
inspected so far.

Next, SourceExtractor \citep{Bertin+96} is used to create source
catalogs for use during astrometric and photometric
calibration. Large-scale image distortions are accounted for by using a
distortion model pre-computed using SCAMP \citep{Bertin-scamp} that is assigned
to each frame before source extraction. OTA-wide estimates of the sky-background
are computed by random sampling the sky in small apertures, while avoiding areas
close to detected sources to yield an unbiased sky sample. 

Up to this point, all 13 OTAs have been reduced in parallel, and besides the
actual data frames we have a large number of sky-samples, fringe scalings and
guesses at the telescope pointing error. All this data is now collected to
compute more robust, global sky background values and fringe scaling factors.
Astrometric calibration is achieved by matching the ODI source catalog to 2MASS,
allowing for telescope offsets and rotation as free parameters. Typical WCS
accuracies achieved via this technique are better than $\rm rms \le 0.2$
arcsec. Global fitting of a WCS solution allowing for changes in distortion have
been attempted but did not produce sufficiently reliable results without user
intervention. Finally, photometric calibration is achieved by cross-referencing
with stellar catalogs from SDSS.

QR also includes functionality to remove pupil-ghost images both flat-fields and
science frames. As the position of the pupil-ghost image changes after swapping
filters in and out of the instrument, a technique was devised to compute the
pupil-ghost directly from data taken with different rotator angles, eliminating
all gaps between cells and detectors, making the resulting templates applicable
to all data.


\section{Implementation - parallel python}

Python as programming language was chosen for a number of reasons, the
experience of the author and the availability of important packages for
scientific computation and handling of astronomical data, easy handling of FITS
files, as well as broad support in and beyond the astronomical community being
some of them. It soon became clear that this choice also comes with significant
performance advantages. To further speed up processing and owing to the nature
of ODI as 13 independent detectors the data reduction process is largely
parallelized using python's multiprocessing package. This optimization leads to
processing times for basic de-trending (correction until flat-fielding) of a 13
OTA exposure with data throughputs of $\ge10$ Mpixels/second on a 6-core machine
with fast I/O, including solid state and RAM drives. Memory usage, on the other
hand, was not considered as a limiting factor, and the pipeline in turn requires
$4-8$ GB of RAM, depending on the amount of calibrations to be performed.

\section{Pipeline execution - locally or in the cloud}

To run the pipeline on observed data, WIYN users currently have to download all
raw data, including calibrations, to their machine. Due to the amount of data (a
typical calibration dataset alone has $\sim20$ GB) this might not be feasible
for all users. For this reason we are currently implementing the QR pipeline as
a user-workspace into the ODI Portal, Pipeline, and Archive (PPA), hosted at the
Pervasive Technologies Institute (PTI) at Indiana University (IU; visit
http://www.odi.iu.edu; also
see contributions by A. Gopu \cite{O18_adassxxiii} and M. Young
\cite{O11_adassxxiii} in this proceeding). This user-workspace enables users to
select science frames and calibration data from the ODI archive and run the QR
pipeline on the cloud using computing resources at PTI/IU. The goal is to allow
users to a) fully reduce their data without having to download the raw data; b)
perform basic analysis functions such as creating source catalogs or image
cutouts; and c) only download the data products they need from the portal. This
eliminates both potentially long downloads as well as the requirements to have
the required computing resources, a tremendous advantage for researchers in
particular at smaller or teaching-oriented institutions. As such it resembles,
e.g., the on-the-fly processing of HST data at STScI coupled with the
user-interface of the HST High Level Science Archive, but applied here -- for
the first time -- to data of a ground-based observatory.

\section{Results and summary}
\begin{figure}
\includegraphics[width=0.32\textwidth]{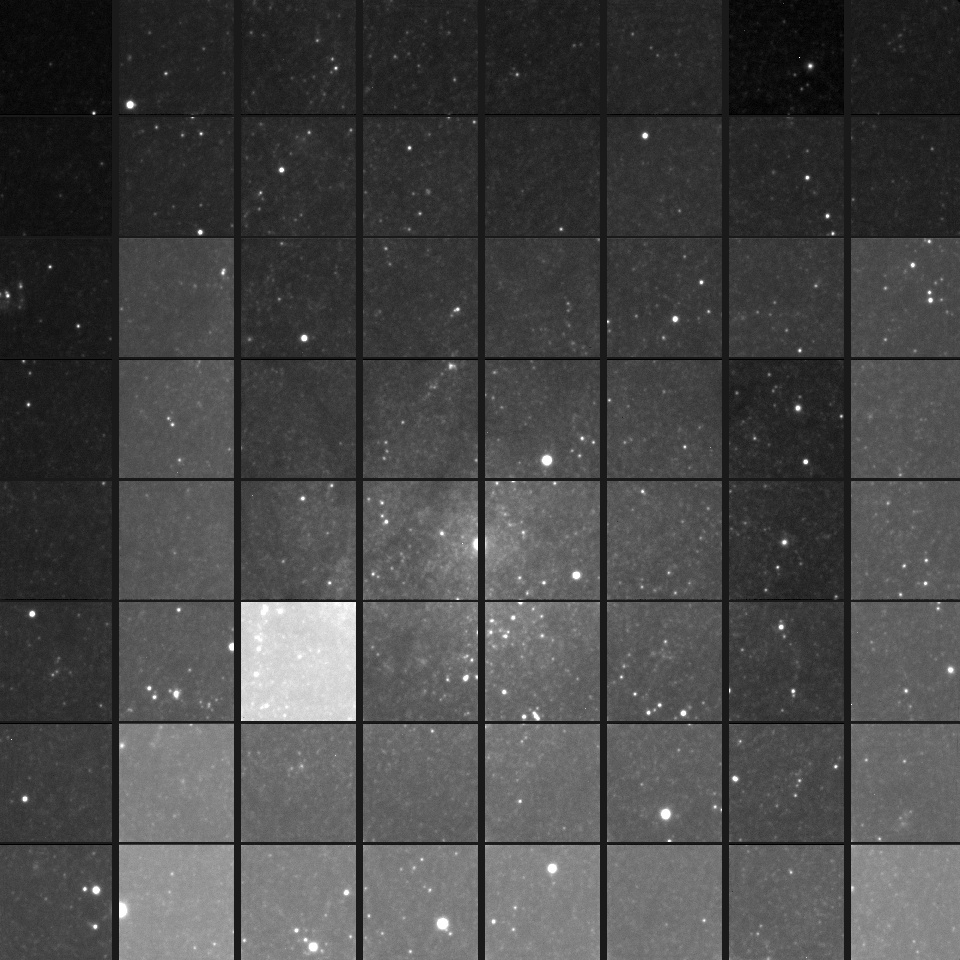}\hfill
\includegraphics[width=0.32\textwidth]{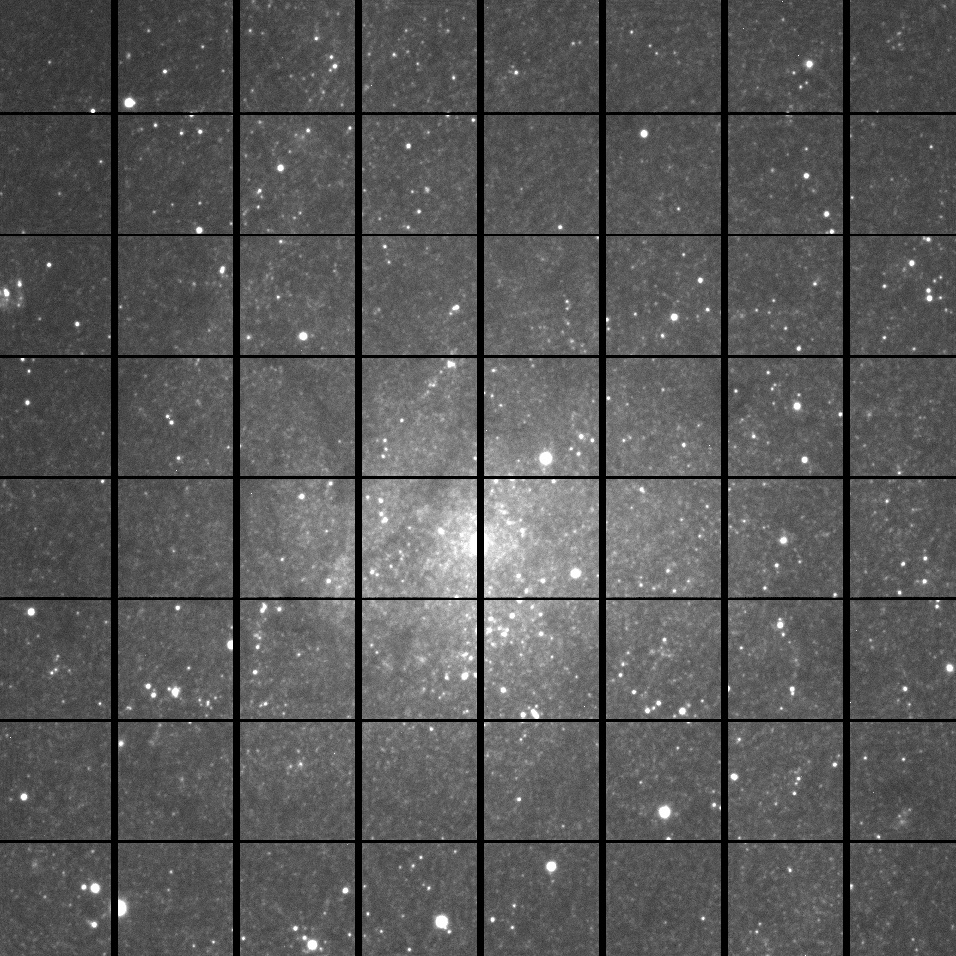}\hfill
\includegraphics[width=0.32\textwidth]{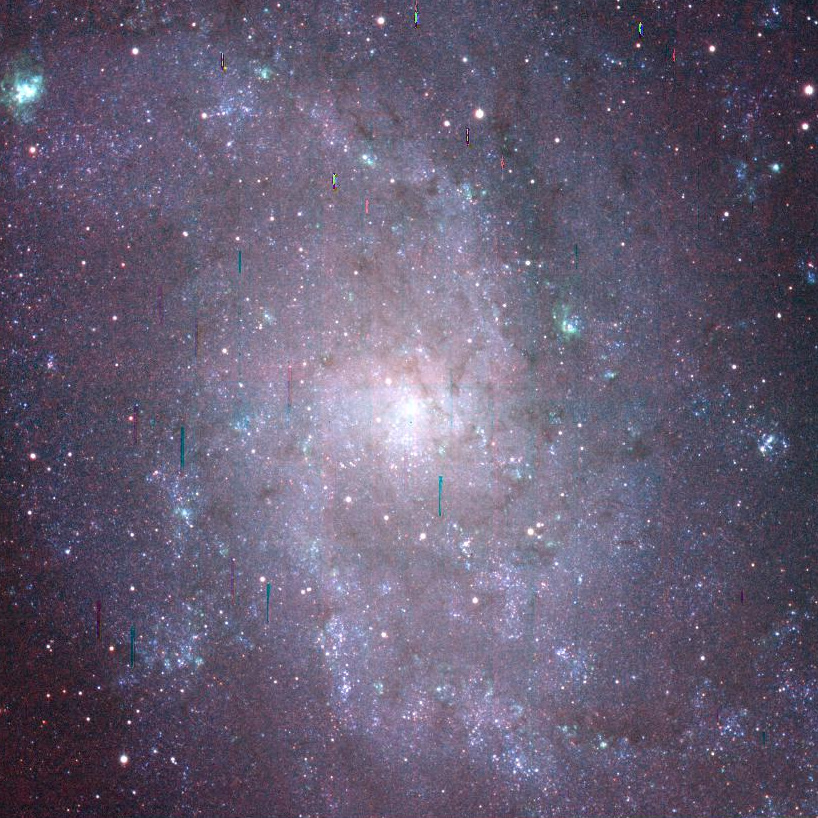}
\caption{Example of M33 before, during and after reduction. Left image: Raw data
of central OTA; Middle panel: Same OTA as on the left, after
flat-fielding. Right panel: RGB image constructed from fully reduced frames in
g,i,z filters.}
\label{fig:p12_f1}
\end{figure}

Shown in Figure \ref{fig:p12_f1} are data products at several stages throughout
the reduction process. Some details still remain to be implemented to improve
the pipeline, but the current performance is very promising and illustrates the
performance of WIYN's newest imager.

In summary, QuickReduce presents a modern approach to a fast and efficient data
reduction for the WIYN One Degree imager as example of modern wide field
imager. It it coded in python with minimal external dependencies and in
particular does not rely on IRAF. The code is parallelized to benefit from
modern multi-core CPUs, achieving reduction times for each $\sim200$ Mpixel frame of $\le20-60$ seconds
depending on the number of reduction steps. It is integrated into the Observers
interface at the telescope, allowing to quickly judge data quality, essential
for efficient observing. It is also integrated into the ODI Portal, Pipeline and Archive
(http://www.odi.iu.edu), giving users access to reduced data without having to download
the significant amounts of raw data generated by ODI. Lastly, QR is publicly
available from the author's website at http://members.galev.org/rkotulla/research/podi-pipeline.

\acknowledgements I thank the organizers of this ADASS XXIII meeting for
financial support. This research was funded by two grants from WIYN Observatory.

\bibliographystyle{asp2010}
\bibliography{P12}


\end{document}